\documentstyle[12pt]{article}

\begin{document}
{\sf \begin{center} \noindent
{\Large \bf Topology of magnetic helicity in torsioned filaments in Hall plasmas}\\[3mm]

by \\[0.3cm]

{\sl L.C. Garcia de Andrade}\\

\vspace{0.5cm} Departamento de F\'{\i}sica
Te\'orica -- IF -- Universidade do Estado do Rio de Janeiro-UERJ\\[-3mm]
Rua S\~ao Francisco Xavier, 524\\[-3mm]
Cep 20550-003, Maracan\~a, Rio de Janeiro, RJ, Brasil\\[-3mm]
Electronic mail address: garcia@dft.if.uerj.br\\[-3mm]
\vspace{2cm} {\bf Abstract}
\end{center}
\paragraph*{}
A solution of magnetic Hall equations for plasma filaments in the
Coulomb gauge is obtained in the non-holonomic frame. Some physical
features of the solution include, the non-conservation of the
magnetic helicity and the decay of the magnetic field in the
filaments. From the mathematical point of view,the presence of
Frenet torsion in the filament is actually shown to be fundamental
for the breaking of conservation of magnetic helicity in the case of
helicoidal filaments. Since the magnetic helicity is not conserved
even in the Coulomb gauge, and the magnetic field decays, one can
say that the dynamo action fails. Actually the presence of torsion
enhances the breaking of magnetic field helicity conservation. A
similar formula of the one obtained here without considering the
Hall effect has been obtained by Moffatt and Ricca (PRSA-1992) in
the case of holonomic filaments. It is shown that unknotted magnetic
filaments may place a lower bound on the magnetic energy.
Discussions on the writhe number are also discussed. \vspace{0.5cm}
\noindent {\bf PACS numbers:}
\hfill\parbox[t]{13.5cm}{02.40.Hw:Riemannian geometries}

\newpage
\newpage
 \section{Introduction}
 Recent interest in the investigation of topology of filaments \cite{1} shows clearly the the investigation of invariants
 such as Calugareanu and magnetic helicity \cite{2,3} are fundamental not only to the knowledge of solar plasma corona
 filaments \cite{4}, but also to filaments used in biology such as the twist of DNA molecules \cite{3}. Investigation of the
 twist and writhe properties of flux tubes, for example would help us to understand the topological properties of filaments in
 solar physics. Another important issues in plasma astrophysics are the turbulence and dynamo action \cite{5}.
 In the case of turbulence , magnetic helicity conservation is fundamental in closed systems, which implies conservation
 in the Coulomb gauge. More recently Kandus \cite{6} has investigated the Hall effect effects on magnetohydrodynamical (MHD)
 turbulence. Earlier Moffatt and Ricca \cite{2} have shown that there is an intimate relation between variation of Frenet torsion
 of the axis of a twisted magnetic flux-tube \cite{7} and the variation of the magnetic helicity, basically showing that
 the magnetic helicity is conserved in the absence of torsion in, what we could call strong criteria and in the weak criteria,
 there would be enough to have that the magnetic helicity would is conserved when torsion does not depend on time, which can
 be a non-zero constant. In this paper we join Kandus idea and Moffatt-Ricca, together by obtaining an expression
 similar to Moffatt-Ricca one in the strong criteria by making use of the Hall effect of imcompressible MHD filament flows
 in non-holonomic Frenet frame. Other interesting features of this filamentary solutions in MHD with Hall effect are that the
 magnetic field decays and there is no energy dissipation of the magnetic energy. Our solution seems to be more realistic
 in the sense that in real plasma , the plasma flow is inhomogeneous and non-holonomy frame would be more suitable for this
 kind of system. The presence of torsion actually enhances the decay of the magnetic field,and as in Moffatt-Ricca
 formula \cite{2} also enhances the breaking of helicity conservation in the
 Coulomb gauge. Some of the values involved in this paper are
 presented in the form of mean-valued physical quantities which can be very useful for future applications in dynamo theory and
 astrophysical turbulence. The paper is organized as
 follows: In section II we consider the MHD equations in the case of non-holonomic magnetic
 filaments. Section III contains the discussion of the
 breaking of conserved quantities of magnetic energy, helicity and the magnetic flux in terms of the torsion of the
 filaments. Section IV presents the Hall equation solution for the torsioned magnetic filaments. Discussions and conclusions are
 presented in section V.
 \section{MHD filamentary torsioned structures}
 In this section the mathematical background is presented. Let us now start by considering the MHD field equations
\begin{equation}
{\nabla}.\vec{B}=0 \label{1}
\end{equation}
\begin{equation}
{\nabla}{\times}{\vec{B}}= {\partial}_{t}\vec{B} \label{2}
\end{equation}
\begin{equation}
{\nabla}.({{\rho}\vec{v}}) +{\partial}_{t}{\rho} =0 \label{3}
\end{equation}
\begin{equation}
{\nabla}{\times}{(\vec{v}{\times}{\vec{B}})} = {\partial}_{t}\vec{B}
\label{4}
\end{equation}
\begin{equation}
\frac{d}{dt}[{\frac{p}{{\rho}^{\gamma}}}]=0 \label{5}
\end{equation}
\begin{equation}
{\rho}\frac{d\vec{v}}{dt}=\vec{J}{\times}{\vec{B}}-{\nabla}p
\label{6}
\end{equation}
where the magnetic quantities involved in these equations are
\begin{equation}
\vec{B}=B_{s}(s,t)\vec{t}+B_{b}(s,t)\vec{b} \label{7}
\end{equation}
\begin{equation}
{\vec{J}}= J(s,t)\vec{t} \label{8}
\end{equation}
\begin{equation}
\vec{u}=u_{s}(s)\vec{t} \label{9}
\end{equation}
The relation between the vector electromagnetic potential and
electric field, which will be fundamental in the investigation of
the Hall effect bellow, is
\begin{equation}
\vec{B}={\nabla}{\times}\vec{A} \label{10}
\end{equation}
where the electric field is
\begin{equation}
\vec{E}=-\frac{{\partial}\vec{A}}{{\partial}t}\label{11}
\end{equation}
in the Coulomb gauge ${\nabla}.\vec{A}=0$. The magnetic field
$\vec{B}$ along the filament is defined by the expression
$\vec{B}=B_{s}(s,t)\vec{t}(s,n,b,t)+B_{b}(s,t)\vec{b}(s,n,b,t)$ and
$B(s,t)$ is the component along the arc length s of the filament
depending upon time. Here we consider that the vectors $\vec{t}$ and
$\vec{n}$ along with binormal vector $\vec{b}$ together form the
Frenet frame which obeys the Frenet-Serret equations
\begin{equation}
\vec{t}'=\kappa\vec{n} \label{12}
\end{equation}
\begin{equation}
\vec{n}'=-\kappa\vec{t}+ {\tau}\vec{b} \label{13}
\end{equation}
\begin{equation}
\vec{b}'=-{\tau}\vec{n} \label{14}
\end{equation}
the dash represents the ordinary derivation with respect to
coordinate s, and $\kappa(s,t)$ is the curvature of the curve where
$\kappa=R^{-1}$. Here ${\tau}$ represents the Frenet torsion. We
follow the assumption that the Frenet frame may depend on other
degrees of freedom such as that the gradient operator becomes
\begin{equation}
{\nabla}=\vec{t}\frac{\partial}{{\partial}s}+\vec{n}\frac{\partial}{{\partial}n}+\vec{b}\frac{\partial}{{\partial}b}
\label{15}
\end{equation}
The electric field is given by
\begin{equation}
\vec{E}=E_{s}\vec{t}+E_{b}\vec{b}\label{16}
\end{equation}
The usual Frenet-Serret equations are
\begin{equation}
\vec{t}'=\kappa\vec{n} \label{17}
\end{equation}
\begin{equation}
\vec{n}'=-\kappa\vec{t}+ {\tau}\vec{b} \label{18}
\end{equation}
\begin{equation}
\vec{b}'=-{\tau}\vec{n} \label{19}
\end{equation}
the dash represents the ordinary derivation with respect to
coordinate s, and $\kappa(s,t)$ is the curvature of the curve where
$\kappa=R^{-1}$. Here ${\tau}$ represents the Frenet torsion. We
follow the assumption that the Frenet frame may depend on other
degrees of freedom such as that the gradient operator becomes
\begin{equation}
{\nabla}=\vec{t}\frac{\partial}{{\partial}s}+\vec{n}\frac{\partial}{{\partial}n}+\vec{b}\frac{\partial}{{\partial}b}
\label{20}
\end{equation}
 The other equations for the other legs of the Frenet frame are
\begin{equation}
\frac{\partial}{{\partial}n}\vec{t}={\theta}_{ns}\vec{n}+[{\Omega}_{b}+{\tau}]\vec{b}
\label{21}
\end{equation}
\begin{equation}
\frac{\partial}{{\partial}n}\vec{n}=-{\theta}_{ns}\vec{t}-
(div\vec{b})\vec{b} \label{22}
\end{equation}
\begin{equation}
\frac{\partial}{{\partial}n}\vec{b}=
-[{\Omega}_{b}+{\tau}]\vec{t}-(div{\vec{b}})\vec{n}\label{23}
\end{equation}
\begin{equation}
\frac{\partial}{{\partial}b}\vec{t}={\theta}_{bs}\vec{b}-[{\Omega}_{n}+{\tau}]\vec{n}
\label{24}
\end{equation}
\begin{equation}
\frac{\partial}{{\partial}b}\vec{n}=[{\Omega}_{n}+{\tau}]\vec{t}-
\kappa+(div\vec{n})\vec{b} \label{25}
\end{equation}
\begin{equation}
\frac{\partial}{{\partial}b}\vec{b}=
-{\theta}_{bs}\vec{t}-[\kappa+(div{\vec{n}})]\vec{n}\label{26}
\end{equation}
The equations \cite{9} for the time evolution of the Frenet frame
yields
\begin{equation}
\dot{\vec{t}}= -{\tau}{\kappa}\vec{n}+ {\kappa}'\vec{b} \label{27}
\end{equation}
\begin{equation}
\dot{\vec{n}}=-{\kappa}{\tau}\vec{t} \label{28}
\end{equation}
\begin{equation}
\dot{\vec{b}}=-{\kappa}'\vec{t} \label{29}
\end{equation}
where ${\kappa}'=\frac{\partial}{{\partial}s}{\kappa}$. The Hall
equation is given by
\begin{equation}
\vec{E}+\vec{u}{\times}\vec{B}=\frac{1}{n_{e}e}({\nabla}{\times}\vec{B}){\times}\vec{B}+\eta{\nabla}{\times}\vec{B}
\label{30}
\end{equation}
where $n_{e}$ is the number density of electrons while e represents
the electric charge. Here $\eta$ represents the resistivity
coefficient. Before we dive into the computations of the Hall
equation for the filaments , let us compute the Coulomb gauge, where
the vector potential is in principle
\begin{equation}
{\vec{A}}= A_{s}(s)\vec{t}+A_{n}(s)\vec{n}+A_{b}(s)\vec{b}
\label{31}
\end{equation}
After a long but straightforward computation one obtains
\begin{equation}
{\partial}_{s}A_{s}+({\theta}_{ns}+{\theta}_{bs})A_{s}=-div{\vec{n}}A_{n}
\label{32}
\end{equation}
where we have considered the following constraints
\begin{equation}
div{\vec{b}}=0\label{33}
\end{equation}
and $A_{b}:=0$. The electric field components can be easily compute
, since
\begin{equation}
\vec{E}=-{\partial}_{t}\vec{A}=-{\partial}_{t}A_{s}\vec{t}-A_{s}{\partial}_{t}\vec{t}
\label{34}
\end{equation}
the last term on the RHS is easily computed if one notes that
\begin{equation}
{\partial}_{t}\vec{t}=\dot{vec{t}}-\vec{u}.{\nabla}\vec{t}
\label{35}
\end{equation}
which by using the above equation for the time total derivative of
$\vec{t}$ yields the components
\begin{equation}
{E_{s}}=-{\partial}_{t}A_{s} \label{36}
\end{equation}
\begin{equation}
{E_{n}}=-{\kappa}(u_{s}+\tau) A_{s} \label{37}
\end{equation}
\begin{equation}
{E_{b}}={\kappa}' A_{s} \label{38}
\end{equation}
where one has considered $A_{n}:=0$ and that $div{\vec{n}}=0$ to
simplify matters. Note that with these simplifications, one can
solve for the vector potential component as
\begin{equation}
A_{s}=A_{0}(t)e^{-\int{({\theta}_{ns}+{\theta}_{bs})ds}} \label{39}
\end{equation}
By making use of the definition of the magnetic field
\begin{equation}
\vec{B}= {\nabla}{\times}\vec{A} \label{40}
\end{equation}
one obtains the components
\begin{equation}
B_{s}=(\tau+{\Omega})A_{s} \label{41}
\end{equation}
By making use of the definition of the magnetic field
\begin{equation}
{B}_{b}={\partial}_{t}A_{s}+{\kappa}A_{s} \label{42}
\end{equation}
while $B_{n}={\Omega}_{n}A_{n}$ vanishes.
\section{Breaking of helicity invariance from torsioned
non-holonomic magnetic filaments} From the mathematical machinery of
previous section one is now able to compute the usual know conserved
quantities of fluid mechanics and MHD. The first one is the magnetic
helicity given by
\begin{equation}
H:=\int{\vec{B}.\vec{A}d^{3}x} \label{43}
\end{equation}
The time evolution of magnetic helicity can be given as in Boozer
\cite{7}
\begin{equation}
\frac{dH}{dt}:=-\int{\vec{E}.\vec{A}d^{3}x} \label{44}
\end{equation}
which yields the following
\begin{equation}
H:={\pi}R^{2}\int{{A_{s}}^{2}ds} \label{45}
\end{equation}
where R is the constant cross-section filaments. This expression for
helicity can be expressed in terms of the mean value of the
potential vector component. These mean values are particularly
useful when one investigate mean-field dynamo theory \cite{8} and
turbulence. Thus in terms of the average value
\begin{equation}
<F(s)>:=\frac{1}{\int{ds}}\int{F(s)ds} \label{46}
\end{equation}
of the arbitrary function $F(s)$, the helicity derivative becomes
\begin{equation}
\frac{dH}{dt}:=\frac{{\tau}_{0}}{2}{\pi}LR^{2}<{\partial}_{t}{A_{s}}^{2}>
\label{47}
\end{equation}
Here $L=\int{ds}$ represents the length of the filament or soar loop
in question. Note that here ${\tau}_{0}$ is the constant torsion of
a helical filament, which is compatible also with $\kappa=\tau$
relation. However imagining the nonconstant torsion case where the
time derivative of the $A_{s}$ varies slowly along the length of the
filament thus
\begin{equation}
\frac{dH}{dt}:={\pi}LR^{2}{\partial}_{t}{A_{s}}^{2}\int{{\tau}ds}
\label{48}
\end{equation}
Expression (\ref{48}) is similar to the Moffatt-Ricca expression
\begin{equation}
\frac{dH}{dt}:={\Phi}^{2}\frac{d\cal{T}}{dt} \label{49}
\end{equation}
where ${\cal{T}}$ is the total torsion given by
\begin{equation}
\int{{\tau}ds}\label{50}
\end{equation}
Here  ${\Phi}$ is the magnetic flux. Let us finally compute the
dissipation \cite{9} of the magnetic energy in the filaments
\begin{equation}
\frac{dE}{dt}=
\int{\vec{B}{\times}(\vec{u}{\times}\vec{B}).d\vec{S}} \label{51}
\end{equation}
which in the case of magnetic filaments becomes
\begin{equation}
\frac{dE}{dt}=
\int{[B_{s}\vec{t}{\times}u_{s}B_{b}\vec{n}].\vec{n}d{S}}=0
\label{52}
\end{equation}
Thus the vanishing of dissipation shows that the non-conservation of
helicity may come only from the torsion itself. The magnetic energy
is easily computed as
\begin{equation}
E= \int{B^{2}dV} \label{53}
\end{equation}
or in the case or magnetic torsioned filaments
\begin{equation}
E= \frac{{\pi}{{\tau}_{0}}^{2}R^{2}L}{2}<{A_{s}}^{2}> \label{54}
\end{equation}
If again the torsion is not constant but the variation of the vector
potential is, one has
\begin{equation}
E= \frac{{\pi}\int{{{\tau}_{0}}^{2}ds}R^{2}L}{2}{A_{s}}^{2}
\label{55}
\end{equation}
which from the Moffatt-Ricca theorem, which states that the for an
unknotted filament the total curvature (or torsion here since they
are equal)
\begin{equation}
\int{{{\tau}}^{2}ds}\ge{\frac{4{\pi}^{2}}{L}}
 \label{56}
\end{equation}
This places a lower limit on the magnetic energy. In the next
section we shall consider the solution of Hall equations for the
magnetic filament. Note that in the case total torsion is constant
in the case of Moffatt-Ricca , the magnetic helicity vanishes, while
in the non-holonomic case this does not happen.
\section{Hall effect for torsioned filaments and the decay of the magnetic field}
The Hall effect vector equation above can be computed term by term
as
\begin{equation}
\vec{u}{\times}\vec{B}= u_{s}A_{s}\vec{n} \label{57}
\end{equation}
A long computation yields
\begin{equation}
{\nabla}{\times}\vec{B}=
-[B_{s}({\Omega}_{b}+{\tau}_{0})+B_{b}({\kappa}+div\vec{n})]\vec{t}+
[B_{b}{\theta}_{bs}+{\partial}_{s}B_{b}+{\tau}_{0}B_{s}]\vec{n}-{\Omega}_{b}B_{b}\vec{b}
\label{58}
\end{equation}
From this expression one obtains
\begin{equation}
[{\nabla}{\times}\vec{B}]{\times}\vec{B}=
-B_{b}[B_{s}({\tau}_{0})+B_{b}({\kappa}+div\vec{n})]\vec{n}
+B_{s}[B_{b}{\theta}_{bs}+{\partial}_{s}B_{b}+{\tau}_{0}B_{s}]\vec{b}-
B_{b}[B_{b}{\theta}_{bs}+{\partial}_{s}B_{b}+{\tau}_{0}B_{s}]\vec{t}
\label{59}
\end{equation}
Collecting all these formulas into the Hall effect equation yields
the three scalar equations
\begin{equation}
{\partial}_{s}B_{b}+{\tau}_{0}B_{s}=0\label{59}
\end{equation}
\begin{equation}
(2u_{s}+{\tau}_{0})A_{s}{\kappa}_{0}=-\frac{1}{n_{e}e}{\tau}_{0}B_{b}[B_{s}+B_{b}]+{\eta}[{\partial}_{s}B_{b}+
{\tau}_{0}B_{s}] \label{61}
\end{equation}
\begin{equation}
-{\partial}_{t}A_{s}=\eta{{\tau}_{0}}^{2}(B_{s}+B_{b})-\frac{1}{n_{e}e}B_{b}[{\partial}_{s}B_{b}+
{\tau}_{0}B_{s}]  \label{62}
\end{equation}
Expression (\ref{59}) reduces the last expression to
\begin{equation}
-{\partial}_{t}A_{s}=2\eta{{\tau}_{0}}^{2}A_{s} \label{63}
\end{equation}
Note that this expression may be easily solve to yield the following
\begin{equation}
A_{s}=A_{1}e^{-2\eta{{\tau}_{0}}^{2}t-\int{({\theta}_{ns}+{\theta}_{sb})ds}}
 \label{64}
\end{equation}
where $A_{1}$ is an integration constant. Substitution into
expressions for the magnetic field components one notes that the
magnetic field
\begin{equation}
B_{s}={\tau}_{0}A_{1}e^{-2\eta{{\tau}_{0}}^{2}t-\int{({\theta}_{ns}+{\theta}_{sb})ds}}
 \label{65}
\end{equation}
which shows clearly that the magnetic field decays and that the
Frenet torsion strongly enhances this decay through the squared term
of torsion in the exponent.

\section{Conclusions}
The topological features of magnetic filaments with Hall effect is
examined. It is shown that in torsioned filaments the role of
torsion is fundamental in the breaking of conserved quantities such
as magnetic helicity for example. It is also shown that the energy
is not dissipated and the flux is also conserved. It seems that in
the non-holonomic frame magnetic helicity is only conserved for
planar solar loops. This work , in a certain extent generalizes
previous work by Moffatt and Ricca \cite{2} where there is also a
link between the time evolutions of helicity and torsion in the case
of holonomic frame. There is a possibility of generalizing the
investigation of Riemannian flux-tube \cite{10} to investigate the
Hall effect in these tubes in the same way as was done recently
\cite{11} with dynamos. The torsioned filaments study undertaken
here can be applied in the recent investigation by Toroek and Kliem
\cite{12} which investigated the kink instability coronal plasma
loops.

\end{document}